\documentclass[11pt,a4paper, onecolumn]{article}
\usepackage[utf8]{inputenc}
\usepackage[english]{babel}
\usepackage{amsmath}
\usepackage{amsfonts}
\usepackage{amssymb}
\usepackage{graphicx}
\usepackage{tabularx}
\usepackage{xcolor}
\usepackage{epstopdf}
\usepackage[left=2cm,right=2cm,top=2cm,bottom=2cm]{geometry}
\usepackage{url}
\usepackage[square,numbers]{natbib} 
\usepackage{authblk}
\usepackage{float}
\usepackage{xr}

\author[1,2]{Christoph Haessig }
\author[3]{Flemming Møller}
\affil[1]{Physics and Physical Chemistry of Foods, Wageningen University, PO Box 17, 6700 AA Wageningen, the Netherlands}
\affil[2]{Current address: Department of Chemical Engineering, KU Leuven, Celestijnenlaan 200J, 3001 Leuven, Belgium}
\affil[3]{International Flavors and Fragrances, Edwin Rahrs Vej 38, 8220 Brabrand, Denmark}
\title{Seeing the unseen: laser speckles as a tool for coagulation tracking}

\begin{document}
	\maketitle
	\begin{center}
			*christoph.haessig@kuleuven.be
	\end{center}

	\begin{abstract}
		The ability to measure protein functionality is critical for the development of plant-based products, particularly with respect to gelation behavior, which is vital for food structure and texture. Small amplitude oscillatory shear tests remain the standard for monitoring protein gelation; however, these methods are costly, time-consuming, and require physical contact with the sample. Laser speckle rheology, an optical-based technique, offers a contactless alternative by assessing rheological properties through speckle pattern fluctuations.
		In this work, we present a simple laser speckle rheology setup, utilizing a diode laser and a digital camera, to monitor rheological changes during the rennet coagulation of milk. We use a viscoelasticity index, derived from a two-dimensional linear correlation, to quantify speckle pattern fluctuations. The laser speckle rheology method is compared with conventional small amplitude oscillatory shear rheology. Results demonstrate that key characteristics of the coagulation process, including coagulation and gelation times, are temporally aligned between the two methods. Furthermore, the viscoelasticity index allows for the comparison of the complex modulus in samples with similar compositions under consistent acquisition parameters.
		These findings underscore the potential of laser speckle rheology as a cost-effective, rapid, and contactless approach for capturing protein gelation, providing a viable alternative to conventional shear rheological methods.

	\end{abstract}
	
	\subsection*{Keywords}
	laser speckle rheology, rennet coagulation, gelation
	
	\section{Introduction}
	The concern over a sustainable food supply, driven by the increasing global population, consumers' flexitarian diet demands, and health and environmental concerns, has led to increased activity in plant protein research \cite{mcclements2022next, abe2023plant}. Of particular interest is the gelation behavior of plant-based proteins, especially in the context of dairy-free product development, due to the importance of proteins for food structure and texture, and consequently, consumer product acceptance \cite{hettiarachchy2013gelling, nieto2016improved, bruckner2019towards}.
	\par
	Therefore, there is a need to quantify the rheological changes that occur during protein gelation. Typically, small amplitude oscillatory shear (SAOS) rheology is used to fully characterize the rheological changes during gelation \cite{tunick2011small, tieu2022effect, cordobes2004rheology}. An optical-based alternative method to shear rheology is the so-called laser speckle rheology \cite{nader2016evaluation}. Illuminating an optically rough surface with coherent light, such as a laser, results in light being scattered on and below the surface \cite{yokoi2014imaging}. The backscattered light undergoes interference, creating a random pattern of bright and dark spots, known as a laser speckle pattern \cite{dainty2013laser, goodman2007speckle, bouyer2010quantum, rivera2020illumination}. However, as the scattering structures are not static but instead undergo Brownian motion, the laser speckle pattern fluctuates \cite{rabal2008dynamic}. Since the mobility of the scatterers depends on the mechanical properties of the surrounding matrix, one can extract the rheological behavior from the fluctuations of the laser speckle pattern by performing temporal cross-correlation of the speckle patterns \cite{weitz1993diffusing}. Laser speckle rheology has been applied in various fields, such as medical diagnostics and tissue characterization \cite{hajjarian2011measurement}. Recently, laser speckle techniques have been used to characterize various food products, including dairy, ice cream, and milk \cite{postnov2018dairy, da2010transient, bello2024speckle}. For instance, Postnov et al. \cite{postnov2018dairy} applied a laser speckle rheology method to evaluate the viscosity properties of dairy products.
	\par
	In contrast to plant-based proteins, dairy proteins and bovine milk, in general, have been extensively studied in the past decades \cite{o2014milk, haug2007bovine, fox2008milk}. Bovine milk is a complex multi-component fluid, mainly consisting of water, lipids, carbohydrates, and proteins, but also containing trace amounts of minerals, vitamins, hormones, and enzymes \cite{o2014milk, haug2007bovine}. The two main protein fractions present in bovine milk are caseins and whey proteins, which constitute around 80\% and 20\%, respectively \cite{fox2008casein}. Caseins are a group of random coil proteins with molecular weights of about $10^7$ -- $10^9$ Da, resulting in diameters of 50 -- 500 nm \cite{fox2008casein}. Naturally, caseins are present in soluble micellar structures. The exterior of the casein micelles is coated with the $\kappa$-casein fraction, characterized by its large carbohydrate moiety, which solubilizes the casein micelle \cite{walstra1990stability, horne2017rennet}. Casein precipitation may be induced through acidification to a pH of 4.6 or the addition of rennet \cite{lucey2017formation, horne2017rennet}. Rennet coagulation is typically divided into a primary and secondary phase \cite{horne2017rennet}. The primary phase describes the enzymatic hydrolysis of the $\kappa$-casein, resulting in the release of the so-called caseinomacropeptide. This, in turn, reduces the repulsion between the casein micelles, leading to their aggregation and subsequently gelation in the secondary phase \cite{horne2017rennet, britten2022rennet}. The rennet coagulation mechanism is affected by various process parameters, including enzyme type and concentration, temperature, pH, and milk composition \cite{horne2017rennet, britten2022rennet}. Therefore, the rennet coagulation of bovine milk can serve as a well-characterized study system to evaluate the ability of laser speckle rheology to track the rheological changes during protein gelation.
	\par
	In this study, we apply a simple frame-to-frame 2-dimensional correlation analysis of laser speckle images to evaluate the rheological changes during rennet coagulation of milk under various process conditions. The results demonstrate the ability of laser speckle rheology to non-invasively capture the rheological transitions during gelation. This opens up the possibility of tracking rheological changes during the gelation of plant-based protein systems, facilitating functionality research and product development.

	\section{Materials and Methods}
	\subsection{Materials}
	Skimmed milk powder (low heat treatment) was provided by International Flavors and Fragrances. The enzymes Marzyme XT 220 PF (440~IMCU/L) and Chymostar (714~IMCU/L) were provided by International Flavors and Fragrances. Calcium chloride dihydrate (technical grade) was purchased from Sigma-Aldrich (St. Louis, USA). All experiments were performed using deionized water (PURELAB flex 1, ELGA LabWater, High Wycombe, UK).
	
	\subsection{Sample preparation} \label{method milk}
	The milk for the coagulation experiments was prepared by mixing milk powder with a 0.009 M calcium chloride solution at a concentration of 60 g/L (referred to as regular protein content) or 120 g/L (referred to as high-protein content). The solution was stirred for 30~min, achieving complete dissolution of the milk powder. Subsequently, the milk was rested for at least 30 min at the coagulation test temperature using a heating chamber at 21$^{\circ}$C to temper the milk to the test temperature. 
	\\
	The coagulants were diluted prior to inoculating the milk samples. Marzyme and Chymostar were diluted 5-fold and 10-fold, respectively, achieving enzyme concentrations of 40~IMCU/L and 71.4~IMCU/L. The diluted enzyme solutions were used for further experiments within 10 min after preparation.

	\subsection{Time sweep}
	The rheological changes upon rennet coagulation of milk were tracked by time sweeps. The time sweeps were performed with a stress-controlled rheometer (Anton-Paar MCR 302) equipped with a concentric cylinder (27 mm; $V_{\mathrm{sample}}$ = 19 mL) geometry. The storage modulus ($G^{\prime}$) and loss modulus ($G^{\prime\prime}$) were measured within a period of 30~min at a shear strain of 1\%, well within the LVR (see figure S2), at a frequency of $\omega$ of 1 Hz. The measurement was performed with a solvent trap to minimize evaporation. All measurements were performed at 21 $^{\circ}$C in triplicate.
	\\
	Our test methodology proceeded as follows. First, 50 mL of milk, prepared according to section~\ref{method milk}, was inoculated at a concentration of 40~IMCU/L or 80~IMCU/L using the diluted coagulants. Secondly, the sample was transferred to the pre-heated measuring geometry, and the time sweep was started after a resting period of 5 min.

	\subsection{Laser speckle rheology}
\begin{figure}[tbh]
		\centering
		\includegraphics[scale=1]{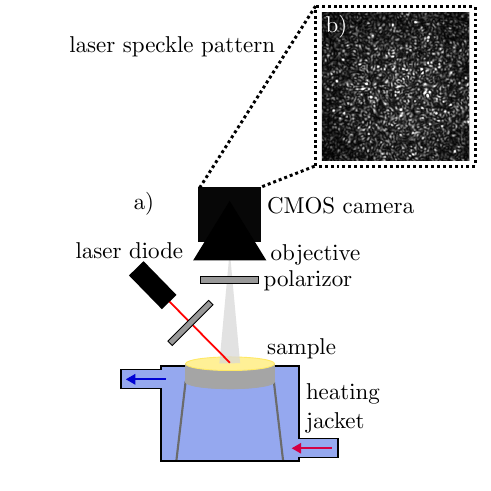}
		\caption{a) Schematic of the  laser speckle rheology set-up and b) laser speckle pattern.}
		\label{fig:laser speckle rheology setup schematic}
	\end{figure}

	The laser speckle rheology optical set-up is schematically displayed in figure \ref{fig:laser speckle rheology setup schematic}. Light from a randomly polarized laser diode (650 nm, 5 mW, Global Laser, Abertillery, UK) was polarized and focused onto the sample. The cross-polarized component of the light back-scattered at an angle of approximately 50$^{\circ}$ was collected by a complementary metal-oxide semiconductor (CMOS) camera (BFS-U3-51S5M-C, FLIR Systems, USA), equipped with a TV lens (25 mm, f1.4, Pentax, Rungis, France). The speckle frames were acquired at 199 fps, over an ROI of 200x200 pixels, for approximately 0.5 s unless specified differently. The camera parameters and image acquisition were controlled through a home-written Python-based code. The sample temperature was controlled through a heating jacket with constant water exchange from an external water bath.

	\subsection{Data processing} \label{data processing}
	The temporal evolution of the acquired speckle frames were analyzed by the 2-dimensional linear correlation, 
	\begin{equation} \label{frames correlation}
	C(t, \tau) =                                                                                                                                                                           
			\frac{\sum_{m,n} (I_{m,n}(t) - \bar{I}(t))(I_{m,n}(t + \tau) - \bar{I}(t + \tau))}{\sqrt{\sum_{m,n} (I_{m,n}(t) - \bar{I}(t))^2 \sum_{m,n} (I_{m,n}(t + \tau) - \bar{I}(t + \tau))^2}},
	\end{equation}
	where $\bar{I}(t)$ and $\bar{I}(t+\tau)$ are the mean intensities of the corresponding frames. The frames correlation coefficient $C$ indicates, similar to the Pearson correlation coefficient, the degree of linear correlation between two frames separated by the lag time $\tau$. $C$ is one if the pixel intensities $I$ at $t$ and $t + \tau$ are linearly related, whereas $C$ is zero if the speckle patterns separated by $\tau$ are uncorrelated. As the movement of the scatterers causes random changes in the speckle pattern, the frames correlation coefficient signifies the amount of movement in the sample between two frames \cite{postnov2018dairy}. Contrary to speckle autocorrelation, the frames correlation is less time-consuming, independent of the speckle contrast, and less vulnerable to acquisition noise \cite{postnov2018dairy}, facilitating the simplification of the laser speckle rheology setup. To quantify the frame-to-frame 2-dimensional linear correlation, we introduce the viscoelasticity index $VI_{\tau}$  \cite{postnov2018dairy}:
	\begin{equation} \label{viscoelasticity index}
		VI_{\tau} = \bar{C_{\tau}} \left( \vert  C_{\tau} - M_{\tau}\vert < \vert M_{\tau} - \sigma_{\tau}\vert\right),
	\end{equation}
	where $M_{\tau}$ and $\sigma_{\tau}$  are the median and standard deviation, respectively, for all frames correlation coeficients between frames within a time period of $\Delta t$ separated by $\tau$. The viscoelasticity index is the average of all frames correlation coefficients for a specific $\tau$ within one standard deviation from the median, providing a robust measure of the scatterer mobility \cite{postnov2018dairy}. The viscosity index was calculated as follows: The frames correlation was calculated for all consecutive  frames within a time period of 100 frames according to equation \ref{frames correlation}.  Hence, $\tau$ was set to approximately 5 ms, unless specified differently. Subsequently, the frames correlation coefficients were used to compute he viscoelasticity index accoridning to equation \ref{viscoelasticity index}. To reduce the computational load, the frames correlation coefficients were calculated for the centered 50x50 pixels.
	\par
	To compare laser speckle rheology with small amplitude oscillatory shear rheology  as methods to track the rheological changes upon rennet coagulation of milk, we extract characteristic features from the temporal $VI$ and $G^\ast$ evolution curves. The coagulation times $t_{\mathrm{c},VI}$ and $t_{\mathrm{c},G^{\ast}}$ were determined as the first point deviating more than 500\% or 100\%, respectively, from the moving average. The onset of the $VI$ plateau $t_{\mathrm{2nd \ plt.}}$ was extracted by finding the first point deviating more than 150\% from the reversed moving average. The value of the second $VI$ plateau $VI_{\mathrm{2nd \  plt.}}$ was calculated as the mean of all $VI$ data points following $t_{\mathrm{2nd \ plt.}}$. The steepest $VI$ slope $(\frac{\mathrm{d}VI}{\mathrm{d}t})_{\mathrm{max}}$ was chosen to represent the curd firming rate of the laser speckle rheology and was determined by calculating the slope for each consecutive points. The curd firming rate measured by the SAOS, $\frac{\mathrm{d}G^{\ast}}{\mathrm{d}t}$, was approximated as average slope of the linear $G^\ast$ section after the coagulation point. The cross-over point between $G^{\prime}$ and $G^{\prime\prime}$ was defined as the point at which $G^{\prime}$ becomes larger than $G^{\prime\prime}$.

	\section{Results and Discussion}
	\subsection{Temporal evolution of laser speckle patterns during rennet coagulation}
\begin{figure}[tbh]
		\centering
		\includegraphics[scale=1]{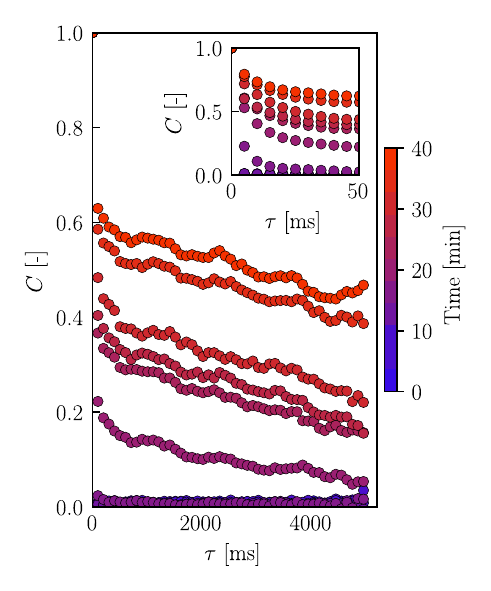}
		\caption{Frames correlation coefficient $C$ as function of the delay time $\tau$ within a $\tau$ range of 5000 ms at various points in time during the rennet coagulation of milk by 40~IMCU/L at 21$^{\circ}$C. The inset shows the initial decay between $\tau$ of 0 to 100 ms for improved visibility.}
		\label{C vs tau}
	\end{figure}

	Upon the rennet addition to milk, chymosin starts to cleave the stabilizing $\kappa$-casein located on the outside of the casein micelles. The cleavage of the $\kappa$-casein leads to the destabilization of the casein micelles. This, in turn, results in the aggregation of the destabilized casein micelles and, eventually, in the formation of a percolating casein gel network \cite{horne2017rennet, britten2022rennet}. The rheology of the milk systems changes drastically as a consequence of this coagulation process, from being a low-viscosity liquid to a firm viscoelastic gel \cite{horne2017rennet, britten2022rennet, karlsson2007RheologicalMilk}. As the scatterers', i.e., casein micelles, mobility is reduced by the formation of aggregates and the subsequently formed space-spanning gel network, we hypothesize that the laser speckle pattern fluctuation slows down simultaneously (see supplementary video). The frames correlation coefficient as a function of the delay time $\tau$ within a range of 5000 ms at various time points during the rennet coagulation by 40~IMCU/L Marzyme at 21$^{\circ}$C is shown in figure \ref{C vs tau}.
	With increasing time after the coagulant addition, the frames correlation coefficient decays slower, as shown in figure \ref{C vs tau}. Within the first twelve minutes, decorrelation of the laser speckle pattern occurs within a single $\tau$ of approximately 5 ms. Thus, within this time frame, the laser speckle fluctuations are sufficiently fast that two subsequent images are uncorrelated at the acquisition frame rate of 199~fps. Thus, the maximum frame rate determines the lower detection limit. On the other extreme, after 40 min, the frames correlation function only decays to approximately 0.5 within 5000~ms. The increased decorrelation time suggests a reduction in the scatterer mobility with increasing time after rennet addition. It is necessary to point out that the frames correlation coefficient is not only affected by the rheological properties but also by the optical properties of the sample \cite{Hajjarian_Nadkarni_2013, tripathi2014assessing}. For instance, Tripathi et al. \cite{tripathi2014assessing} attributed an increase in the temporal speckle intensity autocorrelation at longer lag times to changing optical properties resulting from blood coagulation. During the rennet coagulation process, the casein micelles destabilize, aggregate, and eventually form a space-spanning gel network \cite{horne2017rennet, britten2022rennet}. Consequently, we expect the optical properties of the milk samples to change during the coagulation process. We acknowledge the importance of the optical properties for deriving the viscoelastic moduli, but they are not considered in this study due to the inherent complexity of milk and the focus on comparing the relative rheological changes of compositionally similar samples. In figure S1, we discuss the backscattered light intensity during the coagulation process and its implications on the sample optical properties. Further, we must note the presence of periodic perturbations with a constant frequency of 11~Hz, suggesting a non-random origin. Potential sources of the observed periodic perturbations are discussed in the supplementary information.
	\par
\begin{figure}[tbh]
		\centering	
		\includegraphics[scale=1]{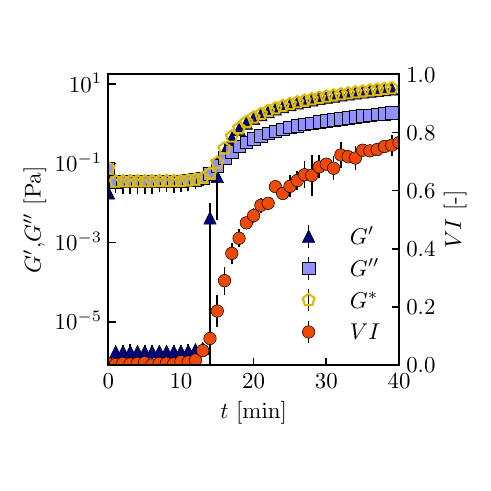}
		\caption{Storage modulus $G^{\prime}$, loss modulus $G^{\prime\prime}$, complex modulus $G^{\ast}$, and viscoelasticity index $VI$ as function of the time after rennet addition $t$ during the rennet coagulation by 40 ~IMCU/L at 21$^{\circ}$C.}
		\label{LSR-SR comparison}
	\end{figure}

	To quantify the linear correlation of the speckle patterns during the coagulation process, thereby providing insights into the simultaneous rheological transitions, we use the viscoelasticity index $VI$. The temporal evolution of the $VI$, as well as the storage modulus $G^{\prime}$, loss modulus $G^{\prime\prime}$, and complex modulus $G^{\ast}$, during the rennet coagulation by 40~IMCU/L at 21$^{\circ}$C is shown in figure \ref{LSR-SR comparison}. Each data point represents the mean of three independent experimental realizations, while the error bars denote one standard deviation.
	SAOS was chosen as the comparison method to laser speckle rheology, as it is the industry standard for detailed rheological characterization and directly probes the mechanical changes upon rennet coagulation \cite{o2002review, lucey2002formation}. Therefore, the comparison of the shear and laser speckle rheology data allows the interpretation of laser speckle rheology data. An oscillatory shear stress or strain is applied, and the resulting response from the developing gel is measured \cite{horne2017rennet}. The applied shear strain must remain within the linear viscoelastic regime during the entire gel development, ensuring a linear proportionality between stress and strain. Yet, initially, the curd is very weak, and consequently, strongly limiting the linear viscoelastic regime. Regardless, we argue that the gel strength develops sufficiently fast, thereby ensuring the linear viscoelasticity. To ensure that the time sweeps were performed within the linear viscoelastic regime, an amplitude sweep was conducted and is shown in figure S2 of the supplementary information.
	\\
	Within the first 13 minutes, both $G^{\prime}$ and $G^{\prime\prime}$ are nearly constant with $G^{\prime} < G^{\prime\prime}$, indicating a dominant liquid-like behavior. After 13 minutes, both $G^{\prime}$ and $G^{\prime\prime}$ increase apparently linearly. Further, a cross-over occurs after approximately 15 minutes, indicating the gel point \cite{ross1995rheological}. To understand the rheological changes upon rennet addition, we will view the system as a developing particle gel with the casein micelles being the particles. For such a system, the shear modulus $G$ can be approximated as \cite{blair1958716}:
	\begin{equation} \label{modulus formula}
		G = C N \frac{\mathrm{d}^2F}{\mathrm{d}x^2},
	\end{equation}
	where $C$ is the characteristic length, $N$ is the number of stress-bearing strands per unit area perpendicular to the direction of the applied deformation $x$, and $\mathrm{d}F$ is the change in free Gibbs energy when particles are moved by $\mathrm{d}x$, i.e., bond strength. According to equation \ref{modulus formula}, during the coagulation process, $G$ depends on both the incorporation of new particles into the gel network and the bond strength. The ratio between the moduli, which remains constant after the cross-over, reflects the nature of the bonds within the network \cite{horne2017rennet}. Consequently, the constant loss factor suggests that the increasing gel firmness is caused by the increasing number of casein micelles being incorporated into the gel network. 
	Similar to the shear moduli, the $VI$ remains constant at approximately zero within the first 13 minutes. Subsequently, the $VI$ increases rapidly before the growth rate starts to continuously decrease at about 16 minutes. After 40 minutes, at the end of the observation window, the $VI$ increased to a maximum value of about 0.75.
		
	The initial $VI$ regime around zero is temporally well aligned with the liquid-like behavior dominated plateau observed for the shear moduli. The dominant liquid-like behavior indicates that the system remains within that time frame in the primary enzymatic phase. The absence of structures resisting the casein micelle mobility within the primary enzymatic phase explains the nearly instant decorrelation of the frames correlation coefficient, responsible for the $VI$ value of zero. 
	The onset of the observed increase in both the moduli and $VI$ at 13 minutes, commonly referred to as the coagulation point \cite{horne2017rennet}, indicates the formation of structures restricting the scatterer mobility, which is simultaneously linked to an increasing resistance to deformation. Thus, this transition implies the onset of the secondary enzymatic phase. With advancing progress in time, the casein aggregates grow further and eventually form a coherent space-spanning gel. The cross-over point was reached after 15.5 minutes, indicating the transition from liquid-like to gel-like behavior, as expected, resulting from the gel formation. Interestingly, the maximum $VI$ growth rate occurred simultaneously at 15.5 minutes. The increasing shear moduli and $VI$ after 15.5 minutes, known as curd firming, are caused by an increasing number of casein micelles incorporated into the particle gel network \cite{horne2017rennet,britten2022rennet,karlsson2007RheologicalMilk}. We propose that the incorporation of casein micelles, and also larger casein aggregates, reduces the number of freely moving casein micelles in solution. Concurrently, the resulting network further hinders the movement of free micelles. Consequently, the number of scatterers that moved between two frames, thereby changing the speckle pattern, characterized by the frames correlation coefficient \cite{postnov2018dairy}, is expected to decrease. Thus, the frames correlation coefficient and consequently the $VI$ increase due to the curd firming.
	\\
	Yet, we must note that the curd firming dynamics captured by the shear moduli and $VI$ differed fundamentally. While the moduli increase visibly linearly (note that figure \ref{LSR-SR comparison} is plotted on a log-linear scale), the slope of the $VI$ seems to decrease with increasing coagulation duration, characteristic of a sigmoidal curve. Generally, one expects that the shear moduli reach a plateau at long times, i.e., a sigmoidal curve, as all casein micelles are incorporated into the gel network \cite{horne2017rennet}. Hence, the SAOS coagulation curve indicates that this last coagulation stage was not reached within the observation window of 40 minutes. Despite the apparent different curd firming dynamics, both the laser speckle rheology and SAOS are temporally well-aligned in terms of the coagulation point and cross-over.

	\subsection{Tracking under various coagulation conditions} 
\begin{figure}[]
		\centering	
		\includegraphics[scale=1]{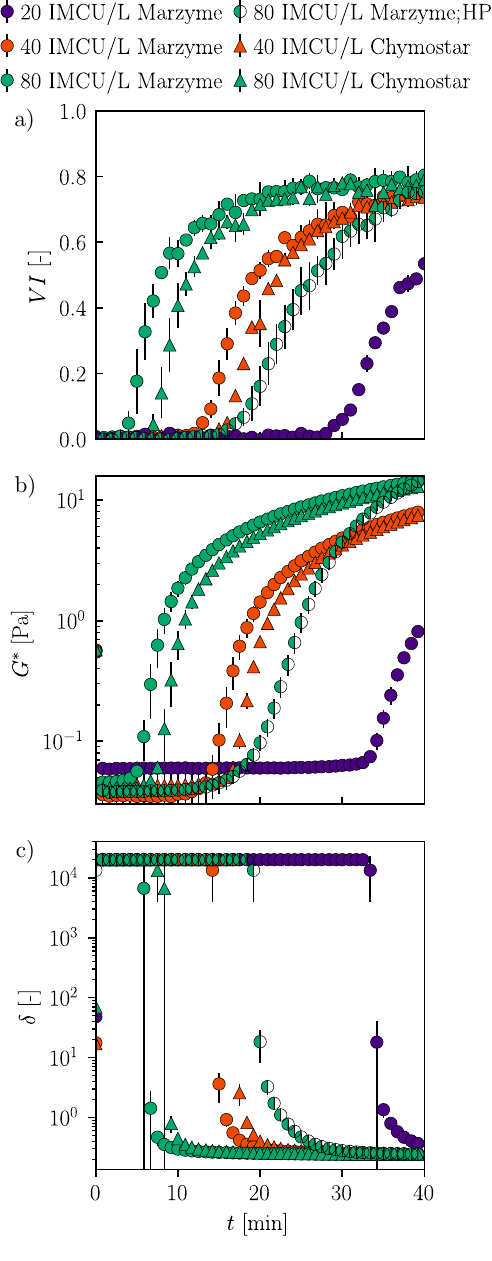}
		\caption{a) Viscoelasticity index $VI$, b) complex modulus $G^{\ast}$, and c) loss factor $\delta$ as a function of time after rennet addition $t$ during the rennet coagulation of regular and high protein (HP) milk by either Marzyme or Chymostar at various concentrations at 21$^{\circ}$C.}
		\label{fig:VI Complex Modulus Rennet Coagulation comparison}
	\end{figure}

\begin{figure*}[htb]
		\centering	
		\includegraphics[scale=1]{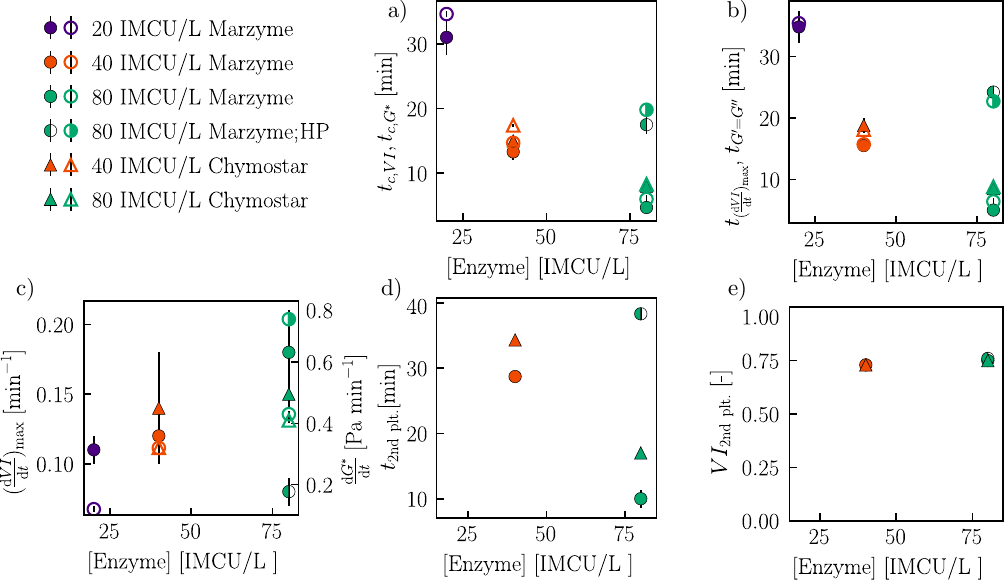}
		\caption{Characteristic features, extracted from the viscoelasticity index and moduli as function of time after rennet addition during the rennet coagulation of regular and high protein (HP) milk by either Marzyme or Chymostar at various concentrations at 21$^{\circ}$C, as function of the enzyme concentration. Open and filled markers are measures extracted from the laser speckle rheology and shear rheometry, respectively. a) coagulation times $t_{\mathrm{c},VI}$ (filled markers) and $t_{\mathrm{c}, G^{\ast}}$ (open markers), b) time of maximum $VI$ slope $t_{(\frac{\mathrm{d}VI}{\mathrm{d}t})_{\mathrm{max}}}$ (filed markers) and cross-over time $t_{G^{\prime}=G^{\prime\prime}}$ (open markers), c) curd firming rates $(\frac{\mathrm{d}VI}{\mathrm{d}t})_{\mathrm{max}}$ (filled markers) and $\frac{\mathrm{d}G^{\ast}}{\mathrm{d}t}$ (open markers) , d) onset of late $VI$ plateau $t_{\mathrm{2nd \ plt.}}$, and e) late $VI$ plateau value $VI_{\mathrm{2nd \  plt.}}$.
			}
		\label{fig:characteristic_features_comparison}
	\end{figure*}

	The generic behavior of both the shear moduli and the viscoelasticity index, and their connection, in response to the rennet coagulation of milk was discussed in the previous section. We further proceed by comparing the response of the two methods to the rennet coagulation of milk under various industry-relevant coagulation conditions. The viscoelasticity index $VI$, complex modulus $G^{\ast}$, and loss factor $\delta$ as a function of time after rennet addition during the rennet coagulation of regular and high-protein, denoted as "HP", milk by various concentrations of Marzyme or Chymostar at 21$^{\circ}$C are shown in figure \ref{fig:VI Complex Modulus Rennet Coagulation comparison}. Here, each data point represents the mean of three independent experimental realizations, while the error bars denote $\pm$ one standard deviation. As visible in figure \ref{fig:VI Complex Modulus Rennet Coagulation comparison}, the rennet coagulation process under all investigated coagulation conditions follows the same stages, as discussed in detail for the rennet coagulation by 40~IMCU/L at 21$^{\circ}$C (see figure \ref{LSR-SR comparison}). Yet, despite the same overall behavior, the temporal occurrence of aggregation, gelation, and subsequent gel firming dynamics differs along the investigated coagulation conditions. To characterize these apparent coagulation dynamics differences, characteristic features were extracted from the coagulation curves shown in figure \ref{fig:VI Complex Modulus Rennet Coagulation comparison} and are presented in figure \ref{fig:characteristic_features_comparison}. A graphical representation of the characteristic features is provided in figures S4 and S5 of the supplementary information. 
	
	The coagulation time is clearly inversely proportional to the enzyme concentration. The coagulation time of the Marzyme-containing samples decreased from approximately 31 minutes to 18 minutes and 5 minutes for enzyme concentrations of 20~IMCU/L, 40~IMCU/L, and 80~IMCU/L, respectively. A similar trend was found for the Chymostar-containing samples, where the coagulation time decreased from around 15 minutes to 8 minutes for enzyme concentrations of 40~IMCU/L and 80~IMCU/L, respectively. Principally, milk coagulation initiates when a sufficiently high degree of $\kappa$-casein is hydrolyzed \cite{mcsweeney2007conversion}. A higher enzyme concentration results in an increased enzymatic activity and consequently reduces the time required to reach the $\kappa$-casein hydrolysis threshold, known as coagulation time. Comparing the two enzymes at a given enzymatic activity, i.e. 40 or 80 IMCU/L, it is noticeable that the Chymostar-containing samples coagulated consistently slightly delayed compared to the Marzyme-containing samples. Despite the theoretical enzymatic activity, experimental conditions like pH and temperature deviating from the standard test definition may have resulted in the different coagulation times. Further, for a fixed enzyme concentration, i.e. 80~IMCU/L Marzyme, doubling the casein fraction resulted in a coagulation time increase from 5 minutes to 18 minutes. This observed coagulation time increase with increasing protein content is due to the reduced enzyme/casein ratio. This, in turn, results in relatively less available enzyme to cleave the $\kappa$-casein. Consequently, the $\kappa$-casein threshold required for coagulation is reached only after longer times \cite{britten2022rennet,karlsson2007RheologicalMilk}.
	Comparing the two methods, we note that the coagulation times determined through SAOS, i.e. $t_{\mathrm{c},G^{\ast}}$, and laser speckle rheology, i.e. $t_{\mathrm{c},VI}$, are approximately the same with margins of about two minutes. As the laser speckle patterns were acquired only every minute, the inter-method coagulation time difference seems negligible, indicating that the two methods are temporally well aligned regarding rheological changes.
	\\
	Similar to the coagulation point, the time of the $G^{\prime}$-$G^{\prime\prime}$ cross-over decreased with increasing enzyme concentration. The increased $\kappa$-casein hydrolysis rate with higher enzyme concentrations results in a shorter enzymatic phase, and thus, an earlier gel point \cite{horne2017rennet}. As discussed for figure \ref{LSR-SR comparison}, the occurrence of the maximum $VI$ growth rate aligns temporally well with the cross-over time for all tested samples, indicating a potential connection between the formation of a space-spanning casein network and the subsequent continuously decreasing $VI$ slope.
	\par
	Further, the development of the gel network was compared by extracting the characteristic curd firming rates. As outlined in section \ref{data processing}, the maximum $VI$ slope and the slope of the linear $G^{\ast}$ section after the coagulation point were extracted for the laser speckle rheology and SAOS, respectively. Again, both methods clearly measured an increase in the curd firming rate with increasing enzyme concentration. The observed proportionality between curd firming rate and enzyme concentration was previously linked to the higher $\kappa$-casein hydrolysis rate \cite{karlsson2007RheologicalMilk}. Both laser speckle rheology and SAOS indicated that the curd firming rate of the Marzyme-containing sample slightly exceeded the Chymostar-containing samples at any investigated enzyme concentration. Yet, despite the consensus regarding the enzyme concentration and type, we note that the two methods indicated an opposing protein content effect on the curd firming rate. The curd firming rate determined by SAOS increased from 0.43 Pa/min to 0.74 Pa/min as a result of a two-fold increase in the casein content during the coagulation by 80~IMCU/L Marzyme. The distance between casein micelles is reduced with increasing casein concentration. Consequently, aggregation between casein micelles is promoted, resulting in the observed increased curd firming rate \cite{mistry2017application}.  In contrast, under the same coagulation conditions, the curd firming rate determined by laser speckle rheology decreased from 0.18 $\mathrm{min}^{-1}$ to 0.08 $\mathrm{min}^{-1}$. We speculate that the apparent decreased firming rate does not reflect the actual gel-firming process. Rather, we hypothesize, as the $VI$ reflects the number of moved scatterers between two frames \cite{postnov2018dairy}, that the elevated number of free casein micelles dominates the increased inclusion of casein into the gel network. 
	\par
	A substantial difference in the apparent coagulation dynamics captured by laser speckle rheology and SAOS is the behavior at the advanced coagulation stage, i.e., as $t$ approached 40 minutes. Within the observation window of 40 minutes, the $G^{\ast}$ increased linearly. Yet, the $VI$ showed sigmoidal behavior with an apparent late-stage $VI$ plateau of approximately 0.75 for all tested conditions. In principle, one expects a sigmoidal $G^{\ast}$ coagulation curve, as eventually all casein micelles are incorporated into the gel network \cite{horne2017rennet}. Yet, as we observed a linear $G^{\ast}$ increase within the experimental timeframe, suggesting that the casein micelles were not depleted yet, the shared late $VI$ plateau value seems to be unrelated to the gel firmness.  Interestingly, the difference between the coagulation time and the onset of the $VI$ plateau seems to decrease with increasing enzyme concentration, except for the high-protein milk sample. The twofold casein concentration increase resulted in a time difference increase from approximately 10 to 21 minutes for the coagulation by 80~IMCU/L.  Rather than being directly related to the gel strength, we hypothesize that the $VI$ plateau is linked to a change in the mobility mode, hence timescale of the scatterer mobility. This timescale transition may be caused by the formation of a space-spanning gel network. Thus, the $\tau$ of 5 ms used for the $VI$ computations were presumably insufficient to capture sufficient scatterer movements to resolve the firmness development during the advanced coagulation stage.

	\subsection{Influence of lag time $\tau$}
\begin{figure*}[tbh]
		\centering
		\includegraphics[scale=1]{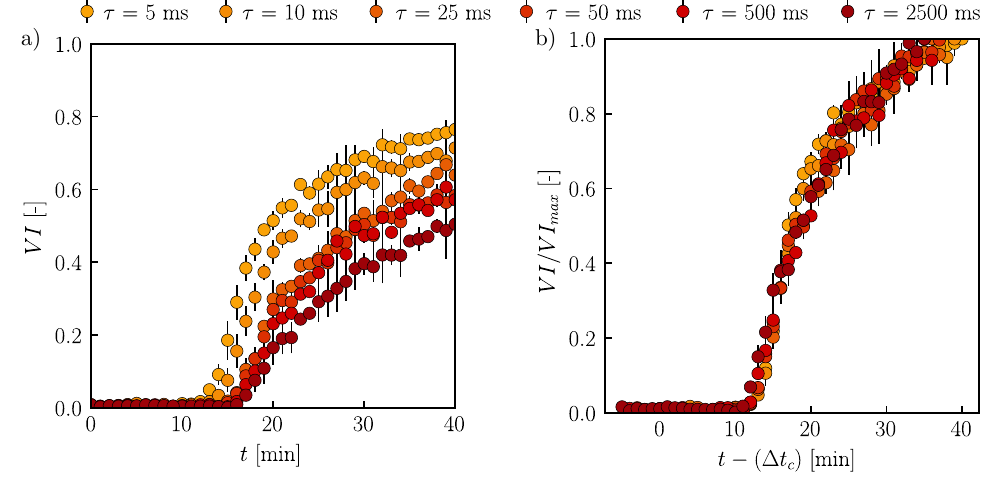}
		\caption{a) Viscoelasticity index $VI$ for lag times 5 ms $\leq \tau \geq$ 2500 ms as a function of time after rennet addition and  b) normalized viscoelasticity index $VI/VI_{max}$ for lag times 5 ms $\leq \tau \geq$ 2500 ms as a function of time after rennet addition, corrected for the difference in the coagulation time relative to $\tau=$ 5 ms, during the rennet coagulation of milk by 40~IMCU/L Marzyme at 21$^{\circ}$C}
		\label{fig:tau effect}
	\end{figure*}

	The lag time $\tau$ is a critical parameter in the laser speckle pattern correlation analysis and must be carefully chosen in respect to the scatterers' timescale \cite{Hajjarian2020TutorialOpportunities}. If $\tau$ is too short ($\tau << \text{Brownian motion}$), insufficient scatterer movement will occur between the two time points, and thus, the laser speckle pattern will appear unchanged. Conversely, if $\tau$ is too long ($\tau >> \text{Brownian motion}$), the laser speckle patterns will appear completely uncorrelated as too many scatterer movements occurred within that timeframe. Hence, in both cases, we are unable to quantify the speed of the laser speckle fluctuations and, consequently, unable to derive any rheological properties.  
	To investigate the impact of the lag time $\tau$ on the viscoelasticity index $VI$ during the rennet coagulation of milk, we determined the $VI$ for various values of $\tau$. The $VI$ for 5 ms $\leq \tau \leq$ 2500 ms as a function of time after rennet addition during the rennet coagulation of milk by 40~IMCU/L Marzyme at 21$^{\circ}$C is shown in figure \ref{fig:tau effect}. Each data point represents the mean of three independent experimental realizations, while the error bars denote $\pm$ one standard deviation.
	The impact of the lag time $\tau$ was quantified by extracting the characteristic features from the $VI$ coagulation curves, as outlined in section \ref{data processing}, and is shown in table \ref{tab:characteristic features tau}. The apparent coagulation time seems to be delayed with increasing $\tau$. The coagulation time increased by 5 minutes, from approximately 13 to 18 minutes, as a result of the $\tau$ increase from 5 ms to 2500 ms. A dependency between $\tau$ and the coagulation time was anticipated. The coagulation point is related to the formation of casein structures that reduce casein mobility. Thus, in turn, fewer scatterer movements occur within a fixed timeframe, leading to the $VI$ increase. Consequently, a longer $\tau$ requires a higher degree of mobility reduction so that the number of scatterer movements is limited sufficiently to avoid complete decorrelation of the two laser speckle patterns. Enhanced mobility reduction is related to a higher degree of structure formation, which requires longer times to develop.
	\par
	Moreover, the onset of the late $VI$ plateau seems to be delayed, while its value decreases, with increasing $\tau$. The delayed onset of the late $VI$ plateau appears to be related to the extended range in which differentiation between scatterer mobility is possible. A larger $\tau$ decreases the $VI$ plateau value due to a greater number of scatterer movements being captured.
	
	Furthermore, we observe that the curd firming rate seems to decrease with increasing $\tau$, potentially related to the reduced $VI$ range with increasing $\tau$.
	\begin{table*}[tbh]
		\centering
		\caption{Coagulation time $t_{\mathrm{c},VI}$, maximum $VI$ slope $(\frac{\mathrm{d}VI}{\mathrm{d}t})_{\mathrm{max}}$, time $t_{\mathrm{2nd \ plt.}}$ and $VI$ value $VI_{\mathrm{2nd \  plt.}}$ of the apparent $VI$ plateau at long times determined for the rennet coagulation of milk by 40~IMCU/L Marzyme at 21$^{\circ}$C at various $\tau$.}
		\vspace{11pt}
		\begin{tabularx}{10cm}{r *{4}{c}}
			\hline
			\multicolumn{1}{c}{$\tau$} & 
			\multicolumn{1}{c}{$t_{\mathrm{c},VI}$} &  
			\multicolumn{1}{c}{$(\frac{\mathrm{d}VI}{\mathrm{d}t})_{\mathrm{max}}$} & 
			\multicolumn{1}{c}{$t_{\mathrm{2nd \ plt.}}$} & 
			\multicolumn{1}{c}{$VI_{\mathrm{2nd \ plt.}}$} \\
			\multicolumn{1}{r}{[ms]} & 
			\multicolumn{1}{c}{[min]} &  
			\multicolumn{1}{c}{[$\mathrm{min}^{-1}$]} & 
			\multicolumn{1}{c}{[min]} & 
			\multicolumn{1}{c}{[-]} \\
			\hline
			5		&	$13.3 \pm 1.2$		&$0.11\pm0.01$ & $34.8\pm2.6$ & $0.73\pm0.01$ \\
			10		&	$15.0\pm0.8$		&$0.12\pm0.03$ & $33.3\pm2.1$ & $0.69\pm0.02$ \\
			25 		&	$16.3\pm0.9$		&$0.11\pm0.02$ & $36.3\pm2.4$ & $0.65\pm0.01$ \\
			50 		&	$16.3\pm0.9$		&$0.12\pm0.03$ & $36.0\pm2.2$ & $0.58\pm0.04$ \\
			500		&	$17.0\pm0.0$		&$0.10\pm0.01$ & $36.0\pm1.4$ & $0.58\pm0.02$ \\
			2500	&	$17.7\pm1.2$		&$0.10\pm0.01$ & $37.7\pm0.5$ & $0.50\pm0.03$ \\
			\hline
		\end{tabularx}
		\label{tab:characteristic features tau}
	\end{table*}
	To compare the underlying coagulation dynamics suggested by the $VI$ coagulation curves at various $\tau$, we normalized $VI$ and $t$. The $VI$ was normalized by dividing it by $VI_{\mathrm{max}}$, removing the differences in the second plateau height and ensuring that the range of the curves is the same for all, i.e., 0–1. Additionally, we introduced the shift factor $\Delta t_c = t_{c,\tau} - t_{c,5\mathrm{ms}}$, which represents the difference in coagulation time relative to the coagulation time of the $VI$ curve at $\tau = 5$ ms. The normalized $VI$ for $5 , \mathrm{ms} \leq \tau \leq 2500 , \mathrm{ms}$ as a function of time after rennet addition during the rennet coagulation of milk by 40 IMCU/L Marzyme at 21 $^{\circ}$C is shown in figure \ref{fig:tau effect} b). Each data point represents the mean of three independent experimental realizations, while the error bars denote $\pm$ one standard deviation.
	Rescaling the $VI$ to the range of 0–1 and temporally aligning the coagulation points results in the collapse of all $VI$ coagulation curves, regardless of $\tau$, into a single master curve. The alignment of the curves through the normalization process reveals that the underlying dynamics captured are inherently similar, regardless of $\tau$. Hence, by adjusting $\tau$, one can extend the measurable range.

	\subsection{Scaling analysis}
\begin{figure}[tbh]
		\centering
		\includegraphics[scale=1]{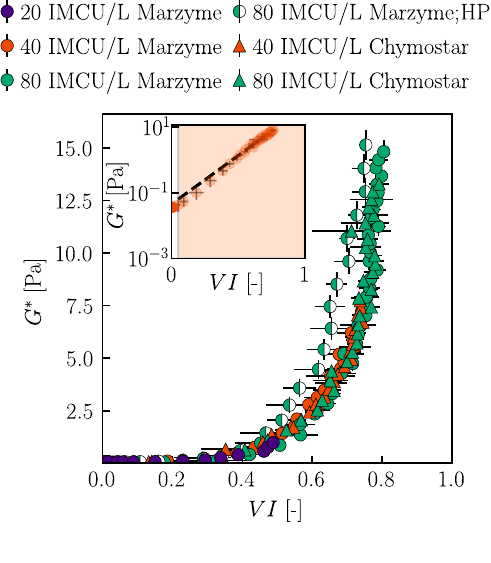}
		\caption{Complex modulus $G^\ast$ as a function of the viscoelasticity index $VI$ for the corresponding time points during the rennet coagulation by variable concentrations of Marzyme or Chymostar at 21$^{\circ}$C. The inset shows an exponential function (dashed line) fitted to the data after the coagulation point, indicate by the light orange shading.}
		\label{fig:scaling analysis}
	\end{figure}

	In the previous sections, we compared the coagulation curves captured by SAOS and laser speckle rheology, and discussed the influence of coagulation conditions on these features. Despite the temporal alignment of key features, such as the coagulation point, and the ability of both methods to capture the effects of various coagulation conditions, we noted that the apparent gel firming dynamics differed between the two techniques. The $G^{\ast}$ evolution suggests linear gel firming within the experimental window, while the $VI$ shows a decreasing firming rate at later stages, characteristic of the sigmoidal $VI$ coagulation curve. To understand the complex relationship between $G^{\ast}$ and $VI$, we plot the complex modulus $G^{\ast}$ as a function of the corresponding viscoelasticity index $VI$ during the rennet coagulation under various coagulation conditions in figure \ref{fig:scaling analysis}. Each data point represents the mean of three independent experimental realizations, with the error bars denoting $\pm$ one standard deviation.
	Interestingly, when plotting $G^{\ast}$ as a function of $VI$, all data collapse onto a single master curve for all coagulation conditions, except for the high protein sample. The collapse of all data from samples with the same composition onto a single master curve indicates that $VI$ can serve as a reliable measure of $G^{\ast}$ under fixed sample composition and acquisition parameters, following appropriate calibration. The deviation of the high-protein content data from the master curve may be related to the increased scatterer fraction resulting from the twofold increase in solids. We speculate that $VI$ is inversely proportional to the amount of freely moving scatterers, namely the non-aggregated casein micelles. Therefore, at higher casein concentrations, a larger fraction of casein must be incorporated into the network to produce the same reduction in speckle activity as observed in the regular milk samples. Consequently, the increased incorporation of casein at higher solid fractions leads to an increase in $G^{\ast}$ at a given $VI$.
	As clearly shown in the inset for coagulation by 40~IMCU/L Marzyme at 21$^{\circ}$C, $G^{\ast}$ appears to be an exponential function of $VI$. The apparent exponential relationship arises from the sigmoidal shape of the $VI$ coagulation curve, with its growth rate decreasing as it approaches the upper limit of $VI = 0.75$, resembling exponential decay. The exponential nature implied by figure \ref{fig:scaling analysis} highlights the sensitivity of the laser speckle rheology method in detecting the initial increase in resistance to deformation caused by casein aggregate formation. However, we note that the differentiation of gel firmness is limited by $VI$ at a constant $\tau$. Thus, laser speckle rheology appears to be a viable method for tracking and differentiating protein aggregation under various conditions, provided the samples have similar compositions and fixed acquisition parameters.
	
	\section{Conclusions}
	In this work, we utilized a simple, cost-effective, and contactless laser speckle rheology setup to monitor the rheological transition upon protein coagulation. Speckle patterns were generated by irradiating milk samples containing 20–80~IMCU/L commercial rennet, which were acquired using a CMOS camera. The speckle patterns were analyzed using a frame-to-frame two-dimensional linear correlation, and particle mobility was extracted using the viscoelasticity index. We compared the ability of the laser speckle rheology setup to monitor the protein coagulation process with the standard small amplitude oscillatory shear (SAOS) rheology.
	We demonstrated the capability of laser speckle rheology to monitor rheological transitions during rennet coagulation of milk under various coagulation conditions. Specifically, critical points in the coagulation process, such as the coagulation and gelation points, were temporally well-aligned between laser speckle rheology and small amplitude oscillatory shear. Notably, the initial increase in resistance to deformation due to aggregate formation was captured with high sensitivity. However, we report an apparent difference in gel firming dynamics between the two methods, which we propose may be due to the viscoelasticity index being linked to scatterer mobility rather than gel firmness itself.
	The effect of coagulation conditions, such as enzyme type, concentration, and protein content, on coagulation dynamics was captured well by laser speckle rheology. Further, we emphasize the importance of the lag time $\tau$ in the frame-to-frame 2-dimensional correlation analysis, as it is essential for matching the scatterer timescale and ensuring accurate temporal alignment.
	These findings highlight the potential of laser speckle rheology as a cost-effective, rapid, and contactless method for assessing protein gelation, offering a promising alternative to conventional rheological techniques.
	
	\subsection*{Abbreviations}
	\textbf{CMOS}: Complementary Metal-Oxide Semiconductor \\
	\textbf{SAOS}: Small Amplitude Oscillatory Shear \\
	\textbf{$VI$}: Viscoelasticity Index
	
	\subsection*{Acknowledgements}
	We thank Birgitte Vesterlund Pedersen and  Michel Hardenberg for their help and suggestions.
	
	\subsection*{Research funding}
	This research was supported by IFF (International Flavors and Fragrances Inc.).
	
	\subsection*{Author contribution}
	\textbf{Christoph Haessig} Conceptualization, Data Curation, Formal Analysis, Investigation, Methodology, Software, Visualization, Writing - original draft
	\\
	\textbf{Flemming Møller} Conceptualization, Funding acquisition, Methodology, Project administration, Resources, Supervision, Writing - review \& editing

	\subsection*{Conflict of interest}
	Flemming Møller is employed at an affiliate of IFF (International Flavors and Fragrances Inc.), but no conflicts of interest arise from this publication. The remaining authors do not have any conflicts of interest to declare.
	
	\subsection*{Data availability statement}
	The datasets generated during and/or analyzed during the current study are available from the corresponding author on reasonable request.
	
	\newpage
	\bibliographystyle{vancouver} 
	\bibliography{LSR_preprint_references} 
	
	\newpage
	\appendix
	\renewcommand{\thefigure}{\Alph{section}\arabic{figure}}  
	\renewcommand{\thefigure}{S\arabic{figure}}
	\setcounter{figure}{0}
	\section{Supplementary information}
	\subsection{Backscattered light intensity}  
	It is expected that the laser speckle fluctuations during the rennet coagulation of milk are affected both by scatterer mobility and the optical properties \cite{Hajjarian_Nadkarni_2013, tripathi2014assessing}. To assess potential changes in optical properties during the rennet coagulation process, we extracted the 90th percentile pixel intensities, $I_{\mathrm{90th}}$, of the speckle patterns as a measure of the backscattering intensity, during the rennet coagulation by 40~{IMCU/L} Marzyme at 21$^{\circ}$C. Specifically, the $I_{\mathrm{90th}}$ was extracted and averaged over 100 speckle patterns per time point. Each data point represents the mean of the 90th percentile, while the error bars denote $\pm$ one standard deviation.
	\begin{figure}[tbh]
		\centering
		\includegraphics[scale=1]{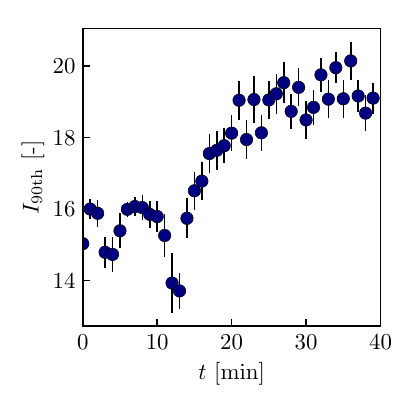}
		\caption{90th percentile pixel intensity $I_{\mathrm{90th}}$ as function of time after rennet addition $t$.}
		\label{fig:light_intensity}
	\end{figure}
	The backscattered light intensity is not constant throughout the rennet coagulation process, as shown in figure \ref{fig:light_intensity}. This variation suggests that the optical properties of the system change during coagulation, likely due to the aggregation of casein micelles, which alters both the scatterer size and number. Interestingly, we observe an increase in pixel intensity around 14 minutes, close to the coagulation time, suggesting that the change in optical properties is closely linked to the coagulation process.

	\subsection{Additional rheological measurements} 
	\label{S:additional rheological measurments}
	A strain sweep was performed to determine the linear viscoelastic regime (LVR) of the rennet-coagulated milk gels. Strain sweeps were carried out using a stress-controlled rheometer (Anton-Paar MCR 302) equipped with a concentric cylinder geometry (27 mm; $V_{\mathrm{sample}}$ = 19 mL). The shear strain $\gamma$ was increased from 0.001 to 1000 at a constant frequency $\omega$ of 1 Hz. All measurements were performed at 32$^{\circ}$C in triplicate.
	
\begin{figure}[tbh]
		\centering
		\includegraphics[scale=1]{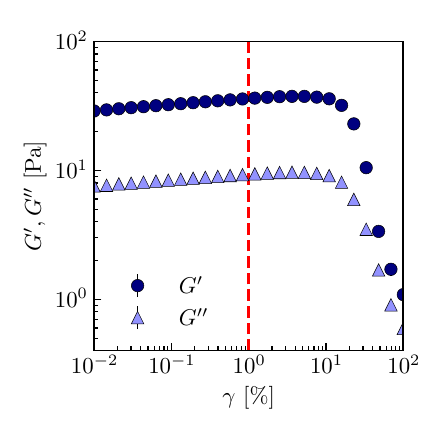}
		\caption{Storage modulus $G^{\prime}$ and loss modulus $G^{\prime\prime}$ as function of the shear amplitude $\gamma$ at a shear frequency of 1 Hz of a milk sample coagulated with 40 IMCU\slash L Marzyme for 30 minutes. The dashed line indicates the shear amplitude used for the time sweeps.}
		\label{fig:amplitude sweep}
	\end{figure}

	Our test methodology proceeded as follows: First, 50 mL of milk, prepared according to section 2.2, was inoculated with a concentration of 40 IMCU/L. Nineteen milliliters of the inoculated milk was transferred to the measuring geometry prior to a gelling period of approximately 1 hour at 32$^{\circ}$C. Subsequently, the strain sweep was performed as described above. The storage modulus $G^{\prime}$ and loss modulus $G^{\prime\prime}$ of a milk sample inoculated with 40 IMCU/L Marzyme at 32$^{\circ}$C after 1 hour are shown in figure \ref{fig:amplitude sweep}. The markers represent the mean of two independent experimental realizations, while the error bars denote $\pm$ one standard deviation.
	As clearly visible in figure \ref{fig:amplitude sweep}, a shear strain of 1\% denoted by the red dashed line, is well within the LVR of the gelled milk system and aligned with previous studies \cite{lucey2000rheological}
	\newpage
	\subsection{Oscillatory noise in frames correlation analysis}
	We must note the presence of periodic perturbations in the frames correlation with a constant frequency of approximately 11 Hz throughout all our measurments. The consistent periodic nature of these perturbations suggests a non-random origin of the noise.
	We propose that the perturbations were introduced by oscillations within the imaging setup, which was not fully vibration-isolated from its surroundings. 

	\newpage
	\subsection{Tracking under various coagulation conditions}
\begin{figure}[tbh]
		\centering
		\includegraphics[scale=1]{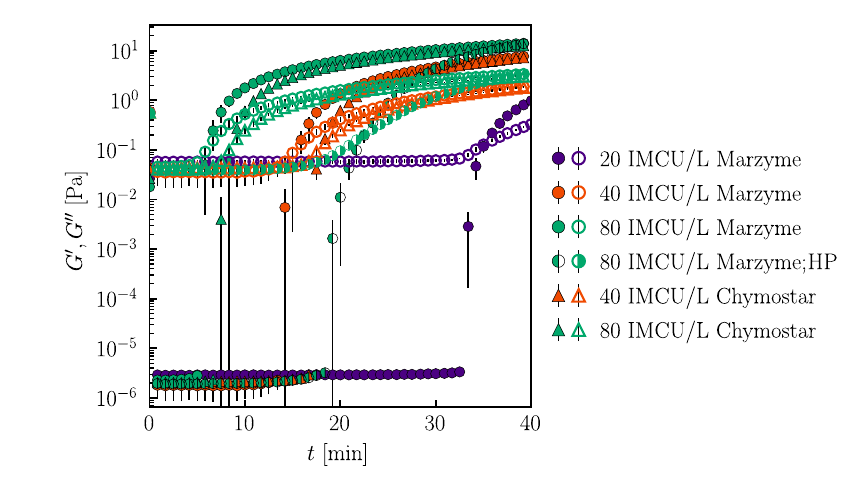}
		\caption{Storage modulus $G^{\prime}$ (filled markers) and loss modulus $G^{\prime\prime}$ (empty markers) as function of time after rennet addition $t$ during the rennet coagulation of regular and high protein (HP) milk by either Marzyme or Chymostar at various concentrations at 21$^{\circ}$C.}
		\label{fig:GG_various}
	\end{figure}

	The storage modulus $G^{\prime}$ and loss modulus $G^{\prime\prime}$ as functions of time after rennet addition during the rennet coagulation of regular and high-protein (denoted as "HP") milk, using various concentrations of Marzyme or Chymostar at 21$^{\circ}$C, are shown in figure \ref{fig:GG_various}. Each data point represents the mean of three independent experimental realizations, with error bars denoting $\pm$ one standard deviation.

	\subsection{Extraction of characteristic features in coagulation curves}
\begin{figure}[h]
		\centering
		\includegraphics[scale=1]{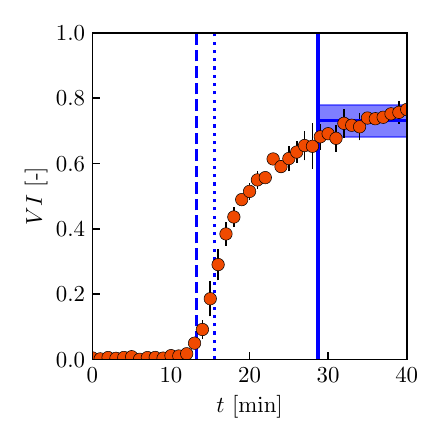}
		\caption{Viscoelastic index $VI$ as function of the time after rennet addition $t$ during the rennet coagulation of milk by 40 IMCU\slash L Marzyme  at 21$^{\circ}$C. Dashed line indicates the coagulation time $t_{\mathrm{c},VI}$. Dotted line indicates the max. $VI$ slope $(\frac{\mathrm{d}VI}{\mathrm{d}t})_{\mathrm{max}}$. Solid line indicates the onset of the late $VI$ plateau $t_{\mathrm{2nd \ plt.}}$ and colored section indicates the points used for calculating the $VI$ of the second plateau $VI_{\mathrm{2nd \  plt.}}$.}
		\label{fig:VI_features}
	\end{figure}

\begin{figure}[H]
		\centering
		\includegraphics[scale=1]{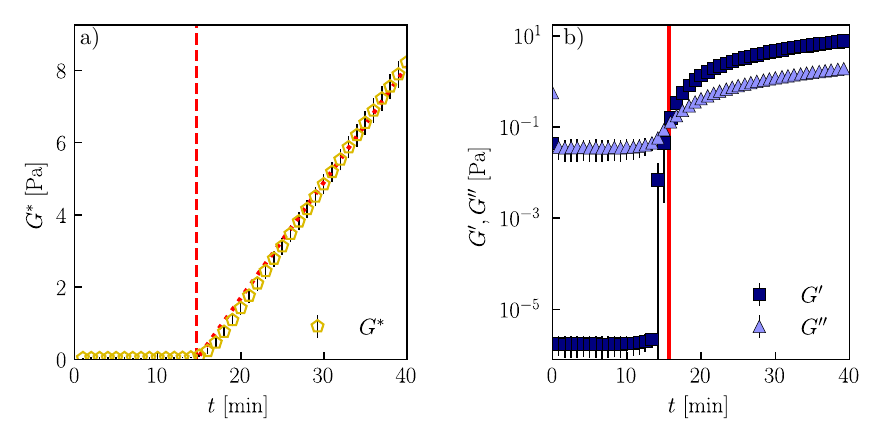}
		\caption{a) Complex modulus $G^{\ast}$ and b) Storage modulus $G^{\prime}$ and loss modulus $G^{\prime\prime}$ as function of time after rennet addition $t$ during the rennet coagulation of milk by 40 IMCU\slash L Marzyme  at 21$^{\circ}$C. Dashed line indicates the coagulation time $t_{\mathrm{c},G^{\ast}}$. Dotted line indciates the slope during curd firming $\frac{\mathrm{d}G^{\ast}}{\mathrm{d}t}$. Solid line indicates the cross-over time $t_{G^{\prime}=G^{\prime\prime}}$. }
		\label{fig:G_features}
	\end{figure}

	The ability of the laser speckle rheology method to monitor rheological transitions during the rennet coagulation of milk was benchmarked against small amplitude oscillatory shear rheology. As outlined in section~2.5, we extracted characteristic features of the coagulation curves to facilitate the benchmarking process. A graphical representation of these characteristic features is provided in figures \ref{fig:VI_features} and \ref{fig:G_features}.
	\\
	In figure \ref{fig:VI_features}, the temporal evolution of the viscoelasticity index ($VI$) during the rennet coagulation by 40 IMCU/L at 21$^{\circ}$C is shown. The markers represent the mean of three independent experimental realizations, with error bars denoting $\pm$ one standard deviation. The blue dashed, dotted, and solid lines indicate the coagulation point, the maximum $VI$ slope, and the onset of the late $VI$ plateau, respectively
	The temporal evolution of the complex modulus ($G^{\ast}$), storage modulus ($G^{\prime}$), and loss modulus ($G^{\prime\prime}$) during the rennet coagulation by 40 IMCU/L at 21$^{\circ}$C is shown in figure \ref{fig:G_features} a) and b), respectively. The markers represent the mean of three independent experimental realizations, with error bars denoting $\pm$ one standard deviation. In figure \ref{fig:G_features} a), we indicate the coagulation time and the slope during curd firming with the red dashed and dotted lines, respectively. The cross-over time is highlighted by the red solid line in figure \ref{fig:G_features} b).
	
\end{document}